\documentclass[aps,twocolumn,showpacs,footinbib]{revtex4}%
\usepackage[latin9]{inputenc}
\usepackage{amssymb}
\usepackage{amsmath}
\usepackage{amscd}
\usepackage{graphicx}
\usepackage{subfigure}
\usepackage{float}
\begin{document}
\draft
\title{Unconventional geometric quantum phase gates with a cavity QED system}
\author{Shi-Biao Zheng\thanks{%
E-mail: sbzheng@pub5.fz.fj.cn}}
\address{Department of Electronic Science and Applied Physics\\
Fuzhou University\\
Fuzhou 350002, P. R. China}
\date{\today }

\begin{abstract}
We propose scheme for realizing two-qubit quantum phase gates via
unconventional geometric phase shift with atoms in a cavity. In the scheme
the atoms interact simultaneously with a highly detuned cavity mode and a
classical field. The atoms undergo no transitions during the gate operation,
while the cavity mode is displaced along a circle in the phase space,
aquiring a geometric phase conditional upon the atomic state. Under certain
conditions, the atoms are disentangled with the cavity mode and thus the
gate is insensitive to both the atomic spontaneous emission and the cavity
decay.
\end{abstract}

 \maketitle \vskip
0.5cm

\narrowtext

\section{INTRODUCTION}

Recently, much attention has been paid to the quantum computers, which are
based on the fundamental quantum mechanical principle. The new type of
machines can solve some problems exponentially faster than the classical
computers [1]. Recent advances in quantum optics have provided powerful
tools for quantum information processing. In cavity QED, schemes have been
proposed for realizing two-qubit quantum logic gates via both the resonant
and dispersive interactions of the atoms with a cavity mode [2]. A quantum
phase gate between a cavity mode and an atom has been demonstrated using
resonant interaction [3]. On the other hand, a scheme has been proposed for
the realization of two-atom entangled states and quantum logic gates within
a nonresonant microwave cavity [4]. The scheme does not require the cavity
mode to act as the memory and the atoms are coupled via the virtual
excitation of the cavity mode. Following the scheme, an experiment has been
reported, in which two Rydberg atoms crossing a nonresonant cavity are
entangled by coherent energy exchange [5].

The above mentioned schemes are based on dynamic evolution. On the other
hand, geometric operation is a promising approach for the implementation of
built-in fault-tolerant quantum phase gates. Compared with the dynamic
gates, the geometric gates may offer practical advantages since the phase is
determined only by the path area, insensitive to the starting state
distributions, the path shape, and the passage rate to traverse the close
path [6]. Thus, geometric phases may be robust against dephasing [7] and the
fidelity of the geometric gates might be significantly higher than that of
the dynamical ones, as demonstrated in a recent experiment in the context of
trapped ions [8]. There are two approaches to obtain the geometric
operations: 1. driving the qubits to undergo appropriate adiabatic cyclic
evolutions; 2. displacing a harmonic oscillator along a closed path
conditional on the state of the qubits. A gate obtained via the first
approach is referred to as a conventional geometric gate, while that
obtained via the second one is referred to as a unconventional geometric
gate [6]. Schemes have been proposed to construct conventional geometric
gates using NMR [9], superconducting nanocircuits [10], and trapped ions
[11]. In comparison with the conventional geometric gates, unconventional
geometric gates does not require additional operations to cancel dynamical
phases and thus simplify experimental operations. The idea of implementing
an unconventional geometric gate has been proposed [8,12] and realized [8]
in a trapped ion system. However, such gates have not been proposed using
other systems.

In this paper we propose a scheme for realizing an unconventional geometric
gate in cavity QED. In the scheme the cavity mode is displaced along a
closed path depending on the atomic states. By this way the system acquires
a phase conditional on the atomic state, producing a phase gate. The atomic
spontaneous emission is suppressed since the atoms undergo no transitions
during the gate operation. Under certain conditions, the displacement
trajectory is a very small circle and thus the cavity mode is disentangled
from the atomic system throughout the operation. In this case the gate is
insensitive to the cavity decay. As far as we know, it is the first scheme
for the implementation of unconventional geometric gates in cavity QED.

This paper is organized as follows. In section 2, we briefly review the
unconventional geometric phase caused by the displacement along a closed
path in phase space. In section 3, we study the evolution for two identical
three-level atoms dispersively interacting with a quantized cavity mode and
a classical field. Section 4 is devoted to the unconventional geometric
phase gates with the cavity QED system. Conclusions appear in section 5.

\section{UNCONVENTIONAL GEOMETRIC PHASE}

We first give a brief review of the geometric phase shift due to
displacement along an arbitrary path [13,14]. The displacement operator is
given by

\begin{equation}
D(\alpha )=e^{\alpha a^{+}-\alpha ^{*}a},
\end{equation}
where a$^{+}$ and a are the creation and annihilation operators of the
harmonic oscillator, respectively. The displacement operators satisfy

\begin{equation}
D(\beta )D(\alpha )=e^{iIm(\beta \alpha ^{*})}D(\alpha +\beta ).
\end{equation}
Consider a path consists of N short straight sections $\Delta \alpha _j$.
Then the total operation is

\begin{eqnarray}
&D_t=D(\Delta \alpha _N)...D(\Delta \alpha
_1)\cr\cr=&e^{iIm\{\sum_{j=2}^N\Delta \alpha
_j\sum_{k=1}^{N-1}\Delta \alpha _k^{*}\}}D(\sum_{j=1}^N\Delta \alpha
_j).
\end{eqnarray}
An arbitrary path $\gamma $ can be approached in the limit
N$\rightarrow \infty $. Thus, we have
\begin{equation}
D_t=D(\int_\gamma d\alpha )e^{i\Theta },
\end{equation}
where
\begin{equation}
\Theta =Im\{\int_\gamma \alpha ^{*}d\alpha \}
\end{equation}
For a closed path we have
\begin{equation}
D_t=D(0)e^{i\Theta }=e^{i\Theta }.
\end{equation}
Set $\alpha =x_1+ix_2$. Then we have

\begin{equation}
\Theta =\oint (x_1dx_2-x_2dx_1).
\end{equation}
Thus, the absolute value of the phase $\Theta $ is equal to two times of the
area involved by the loop in the phase space. The idea has been used to
realize the nonlinear Hamiltonian J$_y^2$ in the context of trapped ions,
where J$_y$ is the collective spin operator [15].

\section{THE CAVITY QED SYSTEM}

We consider two identical three-level atoms, which have one excited state $%
\left| r\right\rangle $ and two ground states $\left| e\right\rangle $ and $%
\left| g\right\rangle $. The quantum information is encoded on the states $%
\left| e\right\rangle $ and $\left| g\right\rangle $ and the state $\left|
r\right\rangle $ is an auxiliary state. The transition $\left|
e\right\rangle \longleftrightarrow \left| r\right\rangle $ is driven by a
cavity mode with the coupling constant g and the detuning $\Delta $ and a
classical laser field with a Rabi frequency $\Omega $ and detuning $\Delta
-\delta $, with $\delta \ll \Delta .$ During the interaction, the state $%
\left| g\right\rangle $ is not affected. In the rotating frame at the cavity
frequency, the Hamiltonian is given by (assuming $\hbar =1$)
\begin{equation}
H_I=\sum_{j=1,2}[\Delta S_{z,j}+(ga^{+}+\Omega e^{i\delta
t})S_j^{-}+(ge^{i\Delta t}a+\Omega e^{-i\delta t})S_j^{+}],
\end{equation}
where $S_{z,j}=\frac 12(\left| r_j\right\rangle \left\langle r_j\right|
-\left| e_j\right\rangle \left\langle e_j\right| )$, $S_j^{+}=\left|
r_j\right\rangle \left\langle e_j\right| $, and $S_j^{-}=\left|
e_j\right\rangle \left\langle r_j\right| $, a$^{+}$ and a are the creation
and annihilation operators for the cavity mode. In the case that $\Delta \gg
\Omega ,g$, the atoms can not exchange energy with the fields. Then the
Hamiltonian of Eq. (1) can be replaced by the effective Hamiltonian, which
includes three parts: 1. the Stark shifts; 2. the dipole coupling between
the two atoms induced by the cavity mode; 3. the coupling between the atoms
and the cavity mode assisted by the classical field. The first two parts has
been derived previously [4]. The third part is characterized by the
transition $\left| r_j,n\right\rangle \longleftrightarrow $ $\left|
r_j,n+1\right\rangle $ and $\left| e_j,n\right\rangle \longleftrightarrow $ $%
\left| e_j,n+1\right\rangle $. The coefficient for $\left|
r_j,n\right\rangle \longrightarrow $ $\left| r_j,n+1\right\rangle $,
mediated by $\left| e_j,n+1\right\rangle $, is given by
\begin{eqnarray}
&&\ \ \frac{\left\langle r_j,n+1\right| H_I\left|
e_j,n+1\right\rangle \left\langle e_j,n+1\right| H_I\left|
r_j,n\right\rangle }\Delta \cr &=&\frac{\left\langle r_j,n+1\right|
\Omega e^{-i\delta t}S_j^{+}\left| e_j,n+1\right\rangle \left\langle
e_j,n+1\right| ga^{+}S_j^{-}\left|
r_j,n\right\rangle }\Delta  \nonumber \\
\ &=&\frac 1\Delta \Omega ge^{-i\delta t}\sqrt{n+1}.
\end{eqnarray}
On the other hand, The effective coupling coefficient for $\left|
e_j,n\right\rangle \longrightarrow $ $\left| e_j,n+1\right\rangle $ is given
by
\begin{eqnarray}
&&\frac{\left\langle e_j,n+1\right| H_I\left| r_j,n\right\rangle
\left\langle r_j,n\right| H_I\left| e_j,n\right\rangle }{-\Delta }
\cr &=&-\frac 1\Delta \Omega ge^{-i\delta t}\sqrt{n+1}.
\end{eqnarray}
Therefore, the effective interaction Hamiltonian is
\begin{eqnarray}
H_i &=&\sum_{j=1,2}\frac 1\Delta [(g^2a^{+}a+\Omega ^2+\Omega
gae^{i\delta t}+\Omega ga^{+}e^{-i\delta t})\cr &&(\left|
r_j\right\rangle \left\langle r_j\right| -\left| e_j\right\rangle
\left\langle e_j\right| )+g^2\left| r_j\right\rangle \left\langle
r_j\right| ]\cr&&+\frac 1\Delta g^2(S_1^{+}S_2^{-}+S_1^{-}S_2^{+}).
\end{eqnarray}
The first two terms in the first smallest bracket describe the Stark shifts
induced by the photons of the cavity field and classical field,
respectively, the other two terms in the first smallest bracket describe the
coupling between the atoms and the cavity mode assisted by the classical
field. The term $\frac 1\Delta g^2\left| r_j\right\rangle \left\langle
r_j\right| $ describes the Stark shift due to the vacuum field. The last two
terms describe the dipole coupling between the two atoms induced by the
cavity mode.

The time evolution of this system is decided by Schr\"odinger's equation:
\begin{equation}
i\frac{d|\psi (t)\rangle }{dt}=H_e|\psi (t)\rangle .
\end{equation}
Perform the unitary transformation
\begin{equation}
|\psi (t)\rangle =e^{-iH_0t}|\psi ^{^{\prime }}(t)\rangle ,
\end{equation}
with
\begin{equation}
H_0=\sum_{j=1,2}\frac 1\Delta [(g^2a^{+}a+\Omega ^2)(\left| r_j\right\rangle
\left\langle r_j\right| -\left| e_j\right\rangle \left\langle e_j\right|
)+g^2\left| r_j\right\rangle \left\langle r_j\right| ].
\end{equation}
Then we obtain
\begin{equation}
i\frac{d|\psi ^{^{\prime }}(t)\rangle }{dt}=H_i^{^{\prime }}|\psi ^{^{\prime
}}(t)\rangle ,
\end{equation}
where
\begin{eqnarray}
H_i^{^{\prime }} &=&\sum_{j=1,2}\frac{\Omega g}\Delta \{[ae^{i(\delta -\frac{%
g^2}\Delta )t}+a^{+}e^{i(-\delta +\frac{g^2}\Delta )t}]\left|
r_j\right\rangle \left\langle r_j\right| \cr
&&\ \ -[ae^{i(\delta +\frac{g^2}\Delta )t}+a^{+}e^{-i(\delta +\frac{g^2}%
\Delta )t}]\left| e_j\right\rangle \left\langle e_j\right|
\}\cr&&+\frac 1\Delta g^2(S_1^{+}S_2^{-}+S_1^{-}S_2^{+}).
\end{eqnarray}

The quantum information is encoded onto the two ground electronic state.
Under the application of $H_i^{^{\prime }}$. The interaction during the
infinitesimal interval [$t$, $t+dt$] induces to the evolution
\begin{eqnarray}
\left| e_1\right\rangle \left| g_2\right\rangle \left| \phi
_{e,g}(t)\right\rangle &\rightarrow &e^{-iH_i^{^{\prime }}dt}\left|
e_1\right\rangle \left| g_2\right\rangle \left| \phi
_{e,g}(t)\right\rangle \cr &=&D(d\alpha )\left| e_1\right\rangle
\left| g_2\right\rangle \left| \phi
_{e,g}(t)\right\rangle ,  \nonumber \\
\left| g_1\right\rangle \left| e_2\right\rangle \left| \phi
_{g,e}(t)\right\rangle &\rightarrow &D(d\alpha )\left| g_1\right\rangle
\left| e_2\right\rangle \left| \phi _{g,e}(t)\right\rangle ,  \nonumber \\
\left| e_1\right\rangle \left| e_2\right\rangle \left| \phi
_{e,e}(t)\right\rangle &\rightarrow &D(2d\alpha )\left| e_1\right\rangle
\left| e_2\right\rangle \left| \phi _{e,e}(t)\right\rangle ,  \nonumber \\
\left| g_1\right\rangle \left| g_2\right\rangle \left| \phi
_{g,g}(t)\right\rangle &\rightarrow &\left| g_1\right\rangle \left|
g_2\right\rangle \left| \phi _{g,g}(t)\right\rangle ,
\end{eqnarray}
where

\begin{equation}
d\alpha =i\frac{\Omega g}\Delta e^{-i(\delta +\frac{g^2}\Delta )t}dt,
\end{equation}
$\left| \phi _{u,v}(t)\right\rangle $ ($u,v=e,g$) denotes the cavity mode
state correlated with the qubit state $\left| u_1\right\rangle \left|
v_2\right\rangle $ at the time $t$. We note that the dipole coupling terms
and the terms containing the population operator $\left| r_j\right\rangle
\left\langle r_j\right| $ have no effect on the evolution since the atoms
are in the ground states, i,e., $(S_1^{+}S_2^{-}+S_1^{-}S_2^{+})\left|
u_1\right\rangle \left| v_2\right\rangle =\left| r_j\right\rangle
\left\langle r_j\right| \otimes \left| u_1\right\rangle \left|
v_2\right\rangle =0$. In fact, only when one atom is in the excited state $%
\left| r_j\right\rangle $ and the other in the ground state $\left|
e_k\right\rangle $ the dipole coupling terms affect evolution of the system.
On the other hand, the population operator $\left| r_j\right\rangle
\left\langle r_j\right| $ only has effect on the state $\left|
r_j\right\rangle $. In the present case, each atom has no probability of
being populated in the excited state $\left| r_j\right\rangle $ and thus we
can ignore the dipole coupling terms and the terms containing the population
operator $\left| r_j\right\rangle \left\langle r_j\right| $.

\section{UNCONVENTIONAL GEOMETRIC PHASE GATES WITH THE CAVITY QED SYSTEM}

Assume that the cavity field is initially in the vacuum state $\left|
0\right\rangle .$ After an interaction time $\tau $ the evolution operators
of the vibrational modes are

\begin{eqnarray}
\left| e_1\right\rangle \left| g_2\right\rangle \left|
0\right\rangle &\rightarrow &e^{i\phi }D(\alpha )\left|
e_1\right\rangle \left| g_2\right\rangle \left| 0\right\rangle , \cr
\left| g_1\right\rangle \left| e_2\right\rangle \left|
0\right\rangle &\rightarrow &e^{i\phi }D(\alpha )\left|
g_1\right\rangle \left|
e_2\right\rangle \left| 0\right\rangle ,  \nonumber \\
\left| e_1\right\rangle \left| e_2\right\rangle \left| 0\right\rangle
&\rightarrow &e^{i\phi ^{^{\prime }}}D(2\alpha )\left| e_1\right\rangle
\left| e_2\right\rangle \left| 0\right\rangle ,  \nonumber \\
\left| g_1\right\rangle \left| g_2\right\rangle \left|
0\right\rangle &\rightarrow &\left| g_1\right\rangle \left|
g_2\right\rangle \left| 0\right\rangle ,
\end{eqnarray}
where
\begin{eqnarray}
\alpha &=&i\int_0^\tau \frac{\Omega g}\Delta e^{-i(\delta
+\frac{g^2}\Delta )t^{^{\prime }}}dt^{^{\prime }} \cr \
&=&-\frac{\Omega g}{\Delta \delta +g^2}[e^{-i(\delta
+\frac{g^2}\Delta )\tau }-1],
\end{eqnarray}
\begin{eqnarray}
\phi &=&Im\int_\gamma \alpha ^{^{\prime }*}d\alpha ^{^{\prime }} \cr
\ &=&-Im\int_0^ti\frac{(\Omega g)^2}{\Delta (\Delta \delta +g^2)}%
(1-e^{-i(\delta +\frac{g^2}\Delta )t^{^{\prime }}})dt^{^{\prime }}
\nonumber \\
\ &=&-\frac{(\Omega g)^2}{\Delta (\Delta \delta +g^2)}[t-\frac 1{(\delta +%
\frac{g^2}\Delta )}\sin (\delta +\frac{g^2}\Delta )t],
\end{eqnarray}
\begin{equation}
\phi ^{^{\prime }}=Im\int_\gamma 4\alpha ^{^{\prime }*}d\alpha ^{^{\prime
}}=4\phi
\end{equation}
Under the condition
\begin{equation}
(\delta +\frac{g^2}\Delta )t=2\pi ,
\end{equation}
the displacement is along a closed path, returning to the original point in
phase space and acquiring a geometric phase conditional upon the electronic
states. This leads to
\begin{eqnarray}
\left| e_1\right\rangle \left| g_2\right\rangle \left|
0\right\rangle &\rightarrow &e^{i\phi }\left| e_1\right\rangle
\left| g_2\right\rangle \left| 0\right\rangle , \cr \left|
g_1\right\rangle \left| e_2\right\rangle \left| 0\right\rangle
&\rightarrow &e^{i\phi }\left| g_1\right\rangle \left|
e_2\right\rangle
\left| 0\right\rangle ,  \nonumber \\
\left| e_1\right\rangle \left| e_2\right\rangle \left| 0\right\rangle
&\rightarrow &e^{i4\phi }\left| e_1\right\rangle \left| e_2\right\rangle
\left| 0\right\rangle ,  \nonumber \\
\left| g_1\right\rangle \left| g_2\right\rangle \left|
0\right\rangle &\rightarrow &\left| g_1\right\rangle \left|
g_2\right\rangle \left| 0\right\rangle ,
\end{eqnarray}
\begin{equation}
\phi =-\frac{(\Omega g)^2}{\Delta (\Delta \delta +g^2)}t.
\end{equation}
Using Eq. (13), we obtain the state evolution for the system governed by $%
H_i $
\begin{eqnarray}
\left| e_1\right\rangle \left| g_2\right\rangle \left|
0\right\rangle &\rightarrow &e^{i(\phi +\Omega ^2t/\Delta )}\left|
e_1\right\rangle \left| g_2\right\rangle \left| 0\right\rangle , \cr
\left| g_1\right\rangle \left| e_2\right\rangle \left|
0\right\rangle &\rightarrow &e^{i(\phi +\Omega ^2t/\Delta )}\left|
g_1\right\rangle \left|
e_2\right\rangle \left| 0\right\rangle ,  \nonumber \\
\left| e_1\right\rangle \left| e_2\right\rangle \left| 0\right\rangle
&\rightarrow &e^{i(4\phi +2\Omega ^2t/\Delta )}\left| e_1\right\rangle
\left| e_2\right\rangle \left| 0\right\rangle ,  \nonumber \\
\left| g_1\right\rangle \left| g_2\right\rangle \left|
0\right\rangle &\rightarrow &\left| g_1\right\rangle \left|
g_2\right\rangle \left| 0\right\rangle .
\end{eqnarray}
After perform the following one-qubit operations

\begin{equation}
\left| e_j\right\rangle \longrightarrow e^{-i(\phi +\Omega ^2t/\Delta
)}\left| e_j\right\rangle ,
\end{equation}
we obtain
\begin{eqnarray}
\left| e_1\right\rangle \left| g_2\right\rangle \left|
0\right\rangle &\rightarrow &\left| e_1\right\rangle \left|
g_2\right\rangle \left| 0\right\rangle , \cr  \left|
g_1\right\rangle \left| e_2\right\rangle \left| 0\right\rangle
&\rightarrow &\left| g_1\right\rangle \left| e_2\right\rangle \left|
0\right\rangle ,  \nonumber \\
\left| e_1\right\rangle \left| e_2\right\rangle \left| 0\right\rangle
&\rightarrow &e^{i2\phi }\left| e_1\right\rangle \left| e_2\right\rangle
\left| 0\right\rangle ,  \nonumber \\
\left| g_1\right\rangle \left| g_2\right\rangle \left|
0\right\rangle &\rightarrow &\left| g_1\right\rangle \left|
g_2\right\rangle \left| 0\right\rangle .
\end{eqnarray}
Choosing
\begin{equation}
2\phi =-\pi ,
\end{equation}
we obtain a $\pi -$phase gate. With the choice $\delta =\frac{g^2}\Delta $, $%
\Omega =g$, $\frac{g^2}\Delta t=\pi $, the conditions (23) and (29) can be
satisfied.

We note that the atoms are entangled with the cavity mode during the gate
operation and thus it is required that the decoherence time of the cavity
mode should be longer than the gate time. However, when $\frac{\Omega g}{%
\Delta \delta +g^2}\ll 1$ the phase-space trajectory is a very small circle
and $D(\alpha )\simeq 1$. In this case the evolution (19) and (21) reduces
to (24) and (25) without the requirement of Eq. (23). Thus, the atoms are
disentangled from the cavity mode throughout the gate operation and the gate
is insensitive to the cavity decay. In this case the phase-space trajectory
should be traversed several times and a longer gate time is required.

We give a brief discussion on the experimental matters. In the scheme the
atoms are never populated in the excited states and thus the decoherence
mainly arises from the cavity decay. Setting $\Delta =10g$, $\delta =2g$,
and $\Omega =g$. Then the required operation time is on the order of $t=\pi
\Delta ^2\delta /(2g^4)\sim 10^2/g$. A cavity with a decay rate $\gamma
=g/27 $ is experimentally achievable [16]. The cavity only has a very small
probability about $\left| \frac{\Omega g}{\Delta \delta +g^2}\right| ^2\sim
10^{-3}$ of being excited during the gate operation. Thus the efficient
decay time of the cavity is about $T=10^3/\gamma \sim 2.7\times 10^4/g$. The
gate error caused by the cavity decay is on the order of $t/T=10^{-2}$, much
smaller than result reported in the Ref. [3].

\section{CONCLUSIONS}

In conclusion, we have proposed a scheme for realizing unconventional
geometric two-qubit phase gates with atoms in a cavity. In the scheme the
atoms interact simultaneously with a highly detuned cavity mode and a
classical field. The atoms remain in their ground states during the gate
operation, while the cavity mode is displaced along a circle in the phase
space, acquiring a geometric phase conditional upon the atomic state. Under
certain conditions, the atomic system is disentangled with the cavity mode
and thus the gate is insensitive to both the atomic spontaneous emission and
the cavity decay.

This work was supported by Fok Ying Tung Education Foundation under Grant
No. 81008, the National Fundamental Research Program Under Grant No.
2001CB309300, the National Natural Science Foundation of China under Grant
Nos. 10225421 and 60008003, and Funds from Fuzhou University.

\end{document}